\documentclass[]{article}
\usepackage{lmodern}
\usepackage{amssymb,amsmath}
\usepackage{ifxetex,ifluatex}
\usepackage{fixltx2e} 
\ifnum 0\ifxetex 1\fi\ifluatex 1\fi=0 
  \usepackage[T1]{fontenc}
  \usepackage[utf8]{inputenc}
\else 
  \ifxetex
    \usepackage{mathspec}
  \else
    \usepackage{fontspec}
  \fi
  \defaultfontfeatures{Ligatures=TeX,Scale=MatchLowercase}
\fi
\IfFileExists{upquote.sty}{\usepackage{upquote}}{}
\IfFileExists{microtype.sty}{%
\usepackage{microtype}
\UseMicrotypeSet[protrusion]{basicmath} 
}{}
\usepackage[margin=1in]{geometry}
\usepackage{hyperref}
\PassOptionsToPackage{usenames,dvipsnames}{color} 
\hypersetup{unicode=true,
            pdftitle={Multilevel Models Allow Modular Specification of What and Where to Regularize, Especially in Small Area Estimation},
            pdfauthor={Michael Tzen},
            colorlinks=true,
            linkcolor=Maroon,
            citecolor=Blue,
            urlcolor=blue,
            breaklinks=true}
\usepackage{longtable,booktabs}
\usepackage{graphicx,grffile}
\makeatletter
\def\maxwidth{\ifdim\Gin@nat@width>\linewidth\linewidth\else\Gin@nat@width\fi}
\def\maxheight{\ifdim\Gin@nat@height>\textheight\textheight\else\Gin@nat@height\fi}
\makeatother
\setkeys{Gin}{width=\maxwidth,height=\maxheight,keepaspectratio}
\IfFileExists{parskip.sty}{%
\usepackage{parskip}
}{
\setlength{\parindent}{0pt}
\setlength{\parskip}{6pt plus 2pt minus 1pt}
}
\providecommand{\tightlist}{%
  \setlength{\itemsep}{0pt}\setlength{\parskip}{0pt}}
\ifx\paragraph\undefined\else
\let\oldparagraph\paragraph
\renewcommand{\paragraph}[1]{\oldparagraph{#1}\mbox{}}
\fi
\ifx\subparagraph\undefined\else
\let\oldsubparagraph\subparagraph
\renewcommand{\subparagraph}[1]{\oldsubparagraph{#1}\mbox{}}
\fi

\let\rmarkdownfootnote\footnote%
\def\footnote{\protect\rmarkdownfootnote}

\usepackage{titling}

  \title{Multilevel Models Allow Modular Specification of What and Where to
Regularize, Especially in Small Area Estimation}
  \pretitle{\vspace{\droptitle}\centering\huge}
  \posttitle{\par}
  \author{Michael Tzen\footnote{California Center for Population Research,
  University of California, Los Angeles}}
  \preauthor{\centering\large\emph}
  \postauthor{\par}
  \predate{\centering\large\emph}
  \postdate{\par}
  \date{May 21, 2018}

\usepackage{amsmath}

\usepackage{amsthm}

\theoremstyle{definition}

\theoremstyle{definition}

\theoremstyle{definition}

\theoremstyle{remark}

\begin{document}
\maketitle

\hypertarget{intro}{%
\section{Intro}\label{intro}}

Through the lens of multilevel model (MLM) specification and
regularization, this is a connect-the-dots introductory summary of Small
Area Estimation, i.e.~small group prediction informed by a complex
sampling design. While a comprehensive book is (Rao and Molina
\protect\hyperlink{ref-rao2015}{2015}), the goal of this paper is to get
interested researchers up to speed with some current developments. We
first provide historical context of two kinds of regularization: 1) the
regularization `within' the components of a predictor and 2) the
regularization `between' outcome and predictor. We focus on the MLM
framework as it allows the analyst to flexibly control the targets of
the regularization. The flexible control is useful when analysts want to
overcome shortcomings in design-based estimates. We'll describe the
precision deficiencies (high variance) typical of design-based estimates
of small groups. We then highlight an interesting MLM example from
(Chaudhuri and Ghosh \protect\hyperlink{ref-chaudhuri2011}{2011}) that
integrates both kinds of regularization (between and within). The key
idea is to use the design-based variance to control the amount of
`between' regularization and prior information to regularize the
components `within' a predictor. The goal is to let the design-based
estimate have authority (when precise) but defer to a model-based
prediction when imprecise. We conclude by discussing optional criteria
to incorporate into a MLM prediction and possible entry points for
extensions.

\hypertarget{rebrand-shrinkage-as-regularization}{%
\section{Rebrand Shrinkage as
Regularization}\label{rebrand-shrinkage-as-regularization}}

As of 2018, there are roughly 91,100 Google Scholar results when
searching with ``shrinkage AND statistic''. There are about 49,900
search results for ``shrinkage AND machine learning''. The boom can be
historically traced to (James and Stein
\protect\hyperlink{ref-james1961}{1961}) decomposing a predictor's Mean
Squared Error, (Efron and Morris
\protect\hyperlink{ref-efron1971}{1971}) evaluating the James-Stein
performance of hierarchical priors (multilevel structure), and
(Tibshirani \protect\hyperlink{ref-tibshirani1996}{1996}) using convex
sparsity constraints to shrink mean components via the lasso. Shrinkage
has become a general tool in statistics and machine learning. Generally
speaking, there are two reasons to use shrinkage:

\begin{enumerate}
\def\labelenumi{\arabic{enumi}.}
\tightlist
\item
  numerical reasons: add suitable constraints in an under-determined
  system to produce a working estimate.
\item
  statistical reasons: shedding off variance by taking on bias
  (incorporating extra information)
\end{enumerate}

The word `shrinkage' gives off the imagery of `shrinking towards zero'.
There are many scenarios where zero is a good value to adjust your
estimates towards (shrink). However, shrinkage does not have to be
towards zero. In other scenarios, shrinkage towards a non-zero quantity
makes sense. For this reason, the rest of this article will primarily
use the term regularize / regularization instead of shrink / shrinkage.

Going forward, we'll use verbiage in an applied regression context
\eqref{eq:regression}. For observation \(i\) we have a left hand side
`outcome' \(y_i\) and a right hand side `predictor' \(f()\) consisting
of `components' \(\beta,X_i,V_i\) and \(A_i\). Specifically, we have a
\(p\)-vector of mean components \(\beta\), for covariates \(X_i\), and
variance components \(V_i\) and \(A_i\). The predicted outcome is
\(\hat y_i\)

\begin{equation}
\hat y_i = f(\beta,X,V_i,A_i)\\
\label{eq:regression}
\end{equation}

We point out that depending on the scenario, some users care more about
the regularized mean-components while others care more about the
regularized outcome. Regularized components may further be combined to
form regularized outcome predictions. Table
\ref{tab:table-regularize-type} categorizes a few scenarios that we will
discuss in later sections.

\begin{longtable}[]{@{}ccccc@{}}
\caption{\label{tab:table-regularize-type} Some Possible Regularization
Scenarios}\tabularnewline
\toprule
Method & Estimand & Default Value & Regularized Towards & Main
Tuner\tabularnewline
\midrule
\endfirsthead
\toprule
Method & Estimand & Default Value & Regularized Towards & Main
Tuner\tabularnewline
\midrule
\endhead
MLM & outcome mean & \(\mu(y_i)^{\text{mle}}\) &
\(f(x_i,\beta,\epsilon_i)\) & \(V_i\)\tabularnewline
Horseshoe & mean component & \(\beta^{\text{mle}}\) & \(0\) &
\(\lambda\)\tabularnewline
Zellner's g & mean component & \(\beta^{\text{mle}}\) &
\(\beta^{\text{prior}}\) & \(g\)\tabularnewline
\bottomrule
\end{longtable}

\hypertarget{reg-comp}{%
\subsection{Regularize Mean Components}\label{reg-comp}}

We first focus on the underlying mean components of a predictor. The
original lasso (Tibshirani
\protect\hyperlink{ref-tibshirani1996}{1996}), regularizes the
parameters for the mean components \(\beta\) towards zero by optimizing
a penalized loss function \eqref{eq:lasso}.

\begin{equation}
\hat\beta^{lasso} = \text{argmin}\{\text{Loss}(y_i,\beta,x_i) + \lambda \left\Vert \beta \right\Vert \}
\label{eq:lasso}
\end{equation}

The original lasso method has a related bayesian version, based on
specifying the `horseshoe prior' for the mean component parameters
(Carvalho, Polson, and Scott
\protect\hyperlink{ref-carvalho2009}{2009}). The horseshoe prior is
below in \eqref{eq:horseshoe}

\begin{equation}
\begin{aligned}
\beta_p & \sim N(0,\lambda_p^2\tau^2) \\
\lambda_p & \sim \text{Half-Cauchy}(0,1) \\
\end{aligned}
\label{eq:horseshoe}
\end{equation}

An alternative to the lasso, is Zellner's g-prior introduced in the
1980s. The g-prior regularizes the MLE \(\beta^{\text{mle}}\) towards an
arbitrary (possibly non-zero) value \(\beta^{\text{prior}}\). (George
and Foster \protect\hyperlink{ref-george2000}{2000}) evaluates Zellner's
g-prior, where we abridge the specification below in \eqref{eq:gprior}.

\begin{equation}
\beta | \psi \sim MVN(\beta^{\text{prior}},g\psi^{-1} (X'X)^{-1})
\label{eq:gprior}
\end{equation}

\hypertarget{reg-out}{%
\subsection{Regularize Outcomes}\label{reg-out}}

As formulated in (C. N. Morris
\protect\hyperlink{ref-morris1983}{1983}), we briefly summarize the
normally distributed MLM in equation \eqref{eq:mlm-morris}. The
MLM-regularized outcomes are \(\theta^*_i\). The observed outcome is
\(y_i\) and \(X_i\) are covariates with unknown mean components
\(\beta\). Unknown variance components \(A_i\) (possibly parameterized)
modify the known variances \(V_i\). Lastly, \(V_i\) and \(A_i\) form
\(R_i\), the regularization factor. The regularization factor controls
how much regularization occurs between the observed outcome \(y_i\) and
the linear predictor \(X_i\beta\). The within group sample size \(n_i\)
plays a role in \(V_i\), which ultimately controls the regularization
factor \(R_i\).

\begin{equation}
\begin{aligned}
y_i | \beta_i,A_i & \sim N(X_i\beta,V_i + A_i) \\
\theta_i | y_i,\beta,A_i & \sim (\theta^*_i,V_i(1-R_i)) \\
\theta^*_i & = (1-R_i)y_i + R_iX_i\beta \\
R_i & = V_i/(V_i+A_i) \\
\end{aligned}
\label{eq:mlm-morris}
\end{equation}

This MLM-type of regularization for outcomes is sometimes referred to as
`borrowing strength' or `partial pooling' across the set of observations
\(\{y_i\}\). This verbiage makes sense if, as often done, an
exchangeability structure is placed on the variance components of the
MLM.

As we'll get to in the following sections. There is a survey statistics
application, called Small Area Estimation, that allows for both
regularization types: regularization ``between'' the outcome and
predictor (`borrowing strength' through the MLM structure) and
``within'' the predictor by regularizing its mean components
(`shrinkage' through prior distributions).

We point out that Small Area Estimation is an overloaded term where some
analysts do not consider incorporating the design variance. In a context
without complex survey weights, a MLM can be applied but will not allow
the design-based variance to control regularization. We are interested
in this design-based variance controlled regularization of a MLM's
prediction. Thus, we next introduce design-based statistics which we
want to improve through a model.

\hypertarget{additional-modeling-of-design-based-statistics}{%
\section{Additional Modeling of Design-Based
Statistics}\label{additional-modeling-of-design-based-statistics}}

There are two frameworks to evaluate the statistical properties of a
computed statistic, the design-based or model-based frameworks. When
computing a statistic \(T\), the frameworks try to pinpoint where the
source of randomness is coming from when taking expectations of a sample
statistic, \(E^{\text{design}}\{T\}\) or \(E^{\text{model}}\{T\}\). This
facilitates concrete statements of mean squared error, bias, variance,
and consistency. (Sterba \protect\hyperlink{ref-sterba2009}{2009})
provides an overview of the different considerations the evaluating
critic must make when choosing a framework scenario. (Skinner and
Wakefield \protect\hyperlink{ref-skinner2017}{2017}) encourages the
analyst to consider hybrid evaluation scenarios where the analyst must
express whether they are interested in: 1) the super population or a
finite population and 2) descriptive estimands or estimands
characterizing relations of a system. For our discussion about
regularization in MLMs, we'll initially start with a design-based
estimate, then show how certain models use the design information to
control regularization. That is, we will look at examples of model-based
predictions of finite population characteristics that are informed by
the sample design (Pfeffermann and others
\protect\hyperlink{ref-pfeffermann2013}{2013}).

In a data set collected using a probability design, say a probability
survey sample \(S\) of individuals \(j\), the analyst typically has
access to the design weights \(w_j^{\text{design}}\). Here we point out
the design weight could solely represent the sample weight
\(w_j^{\text{sample}}\) or it could represent the survey weight
\(w_j^{\text{survey}}\) (where additional adjustments have been made to
the initial sample weight). See equation \eqref{eq:weights} where
\(\pi_j\) is the sample selection probability for individual
\(j \in S\).

\begin{equation}
w_j^{\text{design}} = \begin{cases}
w_j^{\text{sample}} = \frac{1}{\pi_j} & \text{known } \pi_j \\
w_j^{\text{survey}} = h(w_j^{\text{sample}}) & \text{survey adjustments through } h
\end{cases}
\label{eq:weights}
\end{equation}

Going forward, we use \(w_j^{\text{design}}\) the generic terminology
for the design weight. Not only does the analyst need to determine which
scenario is represented by their design weights, but they also
immediately face the question of ``how should I use the design
weights?'' When the analyst's goal is descriptive, say to compute sample
estimates of finite population quantities (totals or proportions), the
design-based framework is usually the de facto option. With that in
mind, we first focus on estimates from the design-based framework but
show how models can improve them. That is, when right hand side
components, possibly from a model, are used to adjust left hand side
outcomes, we only care about the adjusted left hand side outcome and how
the regularization is controlled.\footnote{This article does not assign
  field-specific relational interpretations of parameters, i.e.~causal
  interpretations of right hand side parameters. See (Sterba
  \protect\hyperlink{ref-sterba2009}{2009}) and (Skinner and Wakefield
  \protect\hyperlink{ref-skinner2017}{2017}) for distinctions.}

A primary option for design-based point estimates of a population
quantity \(T\) is to apply the design weights to a weighted estimator,
say the Horvitz-Thompson estimator,
\(\hat T^{\text{HT}} = \sum_{j \in S} \ T_j \frac{1}{\pi_j^{\text{design}}}\).
If the analyst is interested in estimates for a subgroup \(c\) of
individuals \(j\), they can include the dirac indicator function
\(1_{j\in c}\) in their computations, resulting in the subgroup
estimates
\(\hat T^{\text{HT}}_c = \sum_{j \in S} 1_{j\in c} \ T_j \frac{1}{\pi_j^{\text{design}}}\)
.

An important quantity is the variance of the subgroup estimate. Note
that in equation \eqref{eq:variance}, the dirac indicator's role will zero
out certain terms of the variance for units that are not in the subgroup
\(c\).

\begin{equation}
V(\hat T^{\text{HT}}_c) = (1-\frac{n}{N})\frac{n}{n-1}\sum_{j \in S} \{ 1_{j\in c} \ T_j \frac{1}{\pi_j^{\text{design}}}-\frac{1}{n}\hat T^{\text{HT}}_c \}^2
\label{eq:variance}
\end{equation}

Whether the analyst is
\href{http://r-survey.r-forge.r-project.org/survey/example-domain.html}{r-inclined}
or
\href{https://stats.idre.ucla.edu/stata/faq/how-can-i-analyze-a-subpopulation-of-my-survey-data-in-stata/}{stata-inclined},
there are software examples for calculating subgroup estimates. Analysts
can compute the respective variances using taylor-linearization (the
delta-method) \(\hat V_c^{\text{taylor}}(\hat T^{\text{HT}}_c)\), or
using alternatives like the bootstrap or the jacknife. Going forward, we
simplify notation for these design-based estimates of subgroups \(c\) as
\(\hat T_c := \hat T^{\text{HT}}_c\) and
\(\hat V_c := \hat V_c^{\text{taylor}}(\hat T^{\text{HT}}_c)\).

\hypertarget{performance-of-design-based-statistics}{%
\subsection{Performance of Design-Based
Statistics}\label{performance-of-design-based-statistics}}

As the name suggests, a design-based estimate has its uncertainty
primarily attributed to the randomization distribution used in the
sample selection design, see equation \eqref{eq:weights}. These
design-based estimates have been called \textbf{direct
estimates}.\footnote{A term commonly used to highlight their sole
  reliance on a unit's measured data and its design-based probabilities
  of selection} One limitation is that these estimates usually perform
fine at group-level units \(c\) that have `large' within unit sample
size \(n_c\), say states, but the performance degrades if the analyst
looks at more granular group-level units \(c'\) with `small' within unit
sample size \(n_{c'}\), say census tracts.

The cost-quality assessments at various resolution scales are initially
determined by the data provider. For secondary analysis, the analyst may
be highly motivated to produce statistics at more granular groupings
(sub-populations / sub-domains / sub-areas) than what was initially
intended by the designers. An analyst, using modern statistical software
packages,\footnote{Using the \texttt{survey} package in R (Lumley
  \protect\hyperlink{ref-lumley2004}{2004}).} can still use their
desired resolution in defining cell groups \(c\) and compute
design-based estimates \(\hat T_c\) and \(\hat V_c\). However, the
analyst is likely to be underwhelmed at the estimate's high variance
(low precision). In the sub-field of Small Area Estimation, the goal is
to improve the performance of these high resolution group unit
estimates. Many of the Small Area Estimation methods lean on an
underlying model and result in predictions, not estimates, of the
groups.

\hypertarget{mod-sap}{%
\subsection{Models with Design Information to Predict Small Group
Quantities}\label{mod-sap}}

To overcome the precision-shortcomings of direct estimates, (Fay and
Herriot \protect\hyperlink{ref-fay1979}{1979}) used a MLM to improve the
direct estimates. The resulting byproduct of the MLM are
\textbf{design-informed model-based predictions} for small areas, which
we refer to as Small Area Predictions (SAP) going forward.\footnote{They
  have also been called `indirect estimates', to contrast against
  `direct estimates'. This article's rebranding is to clarify that the
  primary MLM output of interest are predictions} The ``small area''
\(c\) are the observational units for the MLM where the SAP predictions
are regularized versions \(\tilde T_c\) of their direct estimate
\(\hat T_c\) counterparts.

The MLM of (Fay and Herriot \protect\hyperlink{ref-fay1979}{1979}) is a
special case of the MLM-class described in (C. N. Morris
\protect\hyperlink{ref-morris1983}{1983}), that we generically described
in \eqref{eq:mlm-morris}. The key application of (Fay and Herriot
\protect\hyperlink{ref-fay1979}{1979}) is that the direct estimates for
unit \(c\) are substituted as inputs \(i \leftarrow c\) into the data
likelihood portion of the MLM, \(y_i \leftarrow \hat T_c\) and
\(V_i \leftarrow \hat V_c\). As discussed in the
\protect\hyperlink{reg-out}{Regularize Outcomes} section, this results
in the MLM regularized prediction \(\theta^*_i \leftarrow \tilde T_c\).
Further, (Fay and Herriot \protect\hyperlink{ref-fay1979}{1979}) used a
non-informative prior for \(\beta\) but no prior for \(A_i\). An
Empirical Bayes procedure produced the resulting predictions. When
viewing this SAP scenario through the language of mixed models, even
though \(V_i\) is usually an estimate, it is sometimes referred to as
the `known' variance of the direct estimate specified in the `sampling
model' while \(A_i\) is a random effect term specified in the `linkage
model' (Rao and Molina \protect\hyperlink{ref-rao2015}{2015})

The MLM regularization trade-off tries to balance how much of the output
prediction \(\tilde T_c\) is composed of the model's linear-predictor
\(x_i\beta\) and how much is composed of the direct design-based
estimate \(\hat T_c\). The variance estimates \(\hat{V_c}\) control the
regularization-factor \(R_c\). For a single small group, it's
MLM-regularized prediction \(\tilde T_c\) should rely more on the direct
estimate when the direct estimate's variance is small, (say if the
sample size within the small group \(c\) is large) but rely more on the
linear-predictor when the variance is large.

After (Fay and Herriot \protect\hyperlink{ref-fay1979}{1979}), many
models to produce SAPs have been proposed, say incorporating spatial
terms for variance components in (Mercer et al.
\protect\hyperlink{ref-mercer2014}{2014}). Researchers have also looked
at alternative likelihoods where (Rabe-Hesketh and Skrondal
\protect\hyperlink{ref-rabe2006}{2006}) uses pseudo likelihoods and
(Chaudhuri and Ghosh \protect\hyperlink{ref-chaudhuri2011}{2011})
considering empirical likelihoods.\footnote{Fay uses the parametric
  normal distribution for the data likelihood. For binary outcomes,
  (Mercer et al. \protect\hyperlink{ref-mercer2014}{2014}) uses the
  asymptotic normal approximation (of the binomial data) for the data
  likelihood. The data likelihood in (Chaudhuri and Ghosh
  \protect\hyperlink{ref-chaudhuri2011}{2011}) is an empirical
  likelihood where exponential family bartlet identities form the moment
  constraints.} For a bayesian treatment, (Chaudhuri and Ghosh
\protect\hyperlink{ref-chaudhuri2011}{2011}) also assessed different
priors. Generally, a MLM structure will produce predictions
\(\tilde T_c\) that are regularized between the direct-estimate
\(\hat T_c\) and the linear-predictor. However, the key property in SAP
models with specific `Fay-Herriot type' regularization is that the
direct-estimate's `known' variance \(\hat V_c\) should play a role in
controlling the amount of regularization.

\hypertarget{mix-and-match-a-mlm-with-both-kinds-of-regularization}{%
\section{Mix and Match: A MLM with Both Kinds of
Regularization}\label{mix-and-match-a-mlm-with-both-kinds-of-regularization}}

In section \ref{mod-sap} we saw (Fay and Herriot
\protect\hyperlink{ref-fay1979}{1979}) apply regularization `between'
the direct estimate outcome and a linear predictor, first described in
section \ref{reg-out}, to produce design-informed predictions for small
groups. Here we additionally introduce regularization `within' the
linear predictor as described in section \ref{reg-comp}. The examples in
(Chaudhuri and Ghosh \protect\hyperlink{ref-chaudhuri2011}{2011})
provide a detailed look at different MLM modeling strategies for
SAP.\footnote{In a single article, (Chaudhuri and Ghosh
  \protect\hyperlink{ref-chaudhuri2011}{2011}) ambitiously evaluated a
  semi-parametric data likelihood along with parametric and
  non-parametric priors for mean components.} For this discussion on
regularization, we focus attention to how (Chaudhuri and Ghosh
\protect\hyperlink{ref-chaudhuri2011}{2011}) studied the behavior of
different priors to regularize the mean components \(\beta\).

If we first start with a MLM structure with a normal likelihood
\eqref{eq:mlm-shrink-2ways-base}, we can set the stage for the desirable
Fay-Herriot type of regularization `between' the direct estimate and a
predictor, controlled by the design-based variance.

\begin{equation}
\begin{aligned}
\hat T_c | \beta,A_c & \sim N(X_c \beta, \ \hat V_c + A_c) \\
R_c & = \hat V_c/(\hat V_c+A_c) \\
\tilde T_c & = (1-R_c) \hat T_c + R_c X_c\beta
\end{aligned}
\label{eq:mlm-shrink-2ways-base}
\end{equation}

To incorporate the second type of `within' predictor regularization of
the mean components \(\beta\), (Chaudhuri and Ghosh
\protect\hyperlink{ref-chaudhuri2011}{2011}) applied the g-prior. As
discussed earlier in the \protect\hyperlink{reg-comp}{Regularize Mean
Components} section, this regularizes \(\beta^{\text{mle}}\) towards
\(\beta^{\text{prior}}\). Therefore, a MLM specified with
\eqref{eq:mlm-shrink-2ways-base} along with \eqref{eq:gprior} will
incorporate both types of regularization. The Fay-Herriot type of
regularization between the direct estimate and the predictor is
controlled by \(R_c\) while the `within' predictor regularization of
mean components is controlled by the hyper parameters \(g\) and
\(\beta^{\text{prior}}\).

Alternatively, we can revisit the horseshoe prior previously described
in \eqref{eq:horseshoe}. If the analyst has many covariates (say many
higher order interactions), regularizing the mean component parameters
towards 0 might be beneficial. Specifying a MLM with
\eqref{eq:mlm-shrink-2ways-base}, to regularize between \(\hat T_c\) and
\(X_c\beta\), with the additional horseshoe prior \eqref{eq:horseshoe}
would result in regularized small group predictions \(\tilde T_c\) whose
underlying mean components \(\beta\) (possibly high dimensional) are
regularized towards 0 by the tuning parameter \(\lambda\).

With all the flexibility a MLM provides, the analyst should first
deliberate how the sample design information is incorporated into the
MLM. Usually, the design weights are used one way or another to estimate
the direct-estimate's variance (as in \eqref{eq:variance}), which are then
used in the MLM as the `known variance' (this was how we presented our
abridged summary in the proceeding discussion). (Fay and Herriot
\protect\hyperlink{ref-fay1979}{1979}) supplied the `known variance' by
backsolving from the coefficient of variation (where design weights were
applied to the quantities of the coefficient of variation). Since
(Mercer et al. \protect\hyperlink{ref-mercer2014}{2014}) modeled the MLM
likelihood with the normal approximation of binomial outcomes, they
backsolved for a binomial effective sample size (starting with scaled
design weights to be used as the normal likelihood's `known variance').
Alternatively, (Rabe-Hesketh and Skrondal
\protect\hyperlink{ref-rabe2006}{2006}) first scaled the design weights
and then incorporated them into a pseudo-likelihood of their MLM. After
that, the analyst should then deliberate if regularizing the components
within a predictor is desireable for their application (say towards 0 or
\(\beta^{\text{prior}}\)).

\hypertarget{discussion}{%
\section{Discussion}\label{discussion}}

We have showcased how specifying a MLM is a way to think generally about
what and where to regularize to. In Small Area Estimation, the goal is
to improve design-based estimates through a model's prediction. Inspired
by an example in (Chaudhuri and Ghosh
\protect\hyperlink{ref-chaudhuri2011}{2011}), we have showcased how a
MLM is a flexible model that can incorporate two kinds of
regularization: `within' the model's predictor mean components and, more
importantly, the Fay-Herriot type of regularization `between' the
predictor and the direct design-based estimates where the regularization
is controlled by the variance of the design-based estimates.

To figure out if a MLM based method is a good approach for the analyst,
we provide a list of questions a consumer of survey data should ask
themselves. Do you have a direct mean and variance estimate that already
incorporates design weights \eqref{eq:weights}? Do you want to improve
group level direct estimates via design-informed model-based
predictions? Do you want your predictions to be regularized between the
design-based estimate and the predictor \eqref{eq:mlm-morris}? Further, do
you want this regularization between the design-based estimate and the
predictor controlled by the design-based variance
\eqref{eq:mlm-shrink-2ways-base}? Do you want to optionally regularize the
components within the predictor \eqref{eq:horseshoe} or \eqref{eq:gprior}?

If the reader thinks model-based predictions of small groups is the
right method for them, (Rao and Molina
\protect\hyperlink{ref-rao2015}{2015}) is an overarching book exposing
the technical details and various extensions for Small Area Estimation.
When considering the extensions, one thing for readers to keep in mind
is that these methods are all about outcome prediction and
regularization. Some extensions incorporate additional variance
component structures, say spatial random effects as in (Mercer et al.
\protect\hyperlink{ref-mercer2014}{2014}). This additional spatial
variance component would add `local' regularization towards neighboring
groups alongiside the `global' regularization towards the predictor.

Along with modeling extensions, (Rao and Molina
\protect\hyperlink{ref-rao2015}{2015}) also discusses two different
topics, that we frame here as loosely related, calibration of design
weights via auxiliary totals and benchmarking of small group estimates.
The loose overlap is the high level motivating criteria; calibration and
benchmarking imposes ``coherence to target quantities''. While there is
an immediate difference in deliveriables, calibration outputs design
weights as a byproduct whereas benchmarking still outputs a (modified)
group prediction, there is also a difference in the type of data source
used for the target quantitites. For example, weight calibration is a
method to produce a set of output calibrated weights\footnote{say, the
  \(h(.)\) adjustment in equation \eqref{eq:weights}} where initially
supplied weights are constrained so that totals using the calibrated
weights are in close agreement with totals from an ideal external data
source,
\(w_i^{\text{calibrate}} \leftarrow \text{calibrate}(w_i^{\text{survey}},x_i^{\text{auxiliary}})\),
see (Särndal \protect\hyperlink{ref-sarndal2010}{2010}). However as
described in (Steorts \protect\hyperlink{ref-steorts2014}{2014}),
``benchmarking a small area estimate'' is a method to constrain the
group-level model-based predictions so that a larger group aggregate of
small group predictions are in close agreement with the direct
design-based estimate of that same larger group,
\(\tilde y_c^{\text{benchmark}} \leftarrow\text{benchmark}(\sum_{c\in L} \tilde y_c\, \sum_{c\in L} \hat y_c)\)
where \(c \in L\) represents that small group \(c\) is an element of the
larger grouping \(L\). Further, calibration uses an external data source
for the targets (say an administrative registry) while benchmarking uses
its own internal data source (the direct design-based estimate).

There are scenarios where researchers advocate supplying small group
totals as auxiliary information into calibration procedures directly, as
in the `GREG Domain Estimation' section in (Rao and Molina
\protect\hyperlink{ref-rao2015}{2015}).\footnote{This is related to an
  alternative framework, model-assisted design-based estimates (Särndal
  \protect\hyperlink{ref-sarndal2010}{2010}). See (Breidt, Opsomer, and
  others \protect\hyperlink{ref-breidt2017}{2017}) for new developments
  and (Skinner and Wakefield \protect\hyperlink{ref-skinner2017}{2017})
  for the design-based or model-based context.} On the other end,
researchers (Steorts and Ugarte
\protect\hyperlink{ref-steorts2014comments}{2014}) have advocated that
external data sources can also be used in benchmarking small group
predictions. At a high-level, benchmarking a MLM prediction to an
external data source is conceptually similar to the Multilevel
Regression and Poststratification (MrP) approach of (Ghitza and Gelman
\protect\hyperlink{ref-ghitza2013}{2013}). However, a fundamental
difference is that MrP simply uses external targets as weights to post
stratify MLM predictions whereas calibration and benchmarking use
constrained matching machinery to impose coherence to their respective
targets.

Discussing this loose overlap between calibration and benchmarking is an
interesting methodlogical note. To not confuse the reader, we describe
the usual ``order of operations'' an analyst may go through when
carrying survey data through the process of producing model-based small
group predictions.

\begin{enumerate}
\def\labelenumi{\arabic{enumi}.}
\setcounter{enumi}{-1}
\tightlist
\item
  Calibration (produce unit level weights)
\end{enumerate}

\begin{itemize}
\tightlist
\item
  commits the initial design weights as primary citizens
\item
  appeals to external authority of auxiliary data to modify initial
  weights
\item
  calibrated weights can immediately be used to produce `design-based'
  estimates and their variances
\end{itemize}

\begin{enumerate}
\def\labelenumi{\arabic{enumi}.}
\tightlist
\item
  Predict Outcomes (MLM incorporating design variance)
\end{enumerate}

\begin{itemize}
\tightlist
\item
  predictor based on covariate features
\item
  the `Fay-Herriot type' of regularization is controlled by the design
  variance
\item
  optionally regularize mean components
\end{itemize}

\begin{enumerate}
\def\labelenumi{\arabic{enumi}.}
\setcounter{enumi}{1}
\tightlist
\item
  Benchmark to Larger Groupings
\end{enumerate}

\begin{itemize}
\tightlist
\item
  the benchmark targets are usually direct estimates (appeal to internal
  authority of large grouping direct estimate)
\item
  optionally incorporate external data source into benchmarking of
  predictions (appeal to external authority of large grouping auxiliary
  data)
\end{itemize}

In special cases, the above order of operations has a nice side effect.
If the design weights are first calibrated and the analyst then fits a
particular MLM for small group predictions, then the outcome predictions
have the nice property of being `self-benchmarked' when using the method
of (You and Rao \protect\hyperlink{ref-you2002}{2002}).

We mentioned the methodological comparison between calibration and
benchmarking so that we can motivate demogrophers to keep up the crucial
work of producing high quality demographic data. As illustrated in all
the examples above, not only are the survey outcomes important, but a
good measure of the underlying population is equally important, if not
more. The second edition of (Rao and Molina
\protect\hyperlink{ref-rao2015}{2015}) had to remove the section on
formal demographic methods.

Formal demographers can apply SAP techniques to their own demographic
application. On the other end, we highlight two entrypoints for
demographers who want to improve SAP methodology. Upstream, in
specifying tailored demographic predictors in the `linking model' (say
non-linear terms motivated by demographic procedures like the Brass
model or the Gompertz model). The challenge would be to incorporate
`symptomatic' demographic approaches into SAP methods. Downstream, in
producing population forecasts to be used as calibration and/or
benchmarking targets (say forecasts of demographic rates as in (Raftery
et al. \protect\hyperlink{ref-raftery2012}{2012})). Whether up or
downstream, demographers have the useful ability to factor in the
``balancing equation'' of population change (more generally, incorporate
knowledge of demographic processes).

Researchers in statistics and machine learning that are developing new
predictive methods should ask if a proposed method allows for the
Fay-Herriot type of regularization controlled by the variance of the
direct design-based estimate. An immediate heuristic to incorporate
predictive methods into a small area prediction system might be to
consider a sequential approach. First, produce a model based prediction
via trees, kernel methods, boosting, ensembling, etc. Then, incorporate
the prediction as a right hand side term in a MLM structure. The second
stage (via the MLM structure) would allow the Fay-Herriot type of
regularization of the direct estimate towards the first-stage
prediction. It would be interesting to develop methods that incorporate
prediction error metrics (from the first stage) as a regularization
controller in the second stage MLM. This heuristic would be a modern
take on `composite estimation' using current predictive techniques for
the `indirect' estimators as described in (Ghosh and Rao
\protect\hyperlink{ref-ghosh1994small}{1994}).

\hypertarget{references}{%
\section*{References}\label{references}}
\addcontentsline{toc}{section}{References}

\hypertarget{refs}{}
\leavevmode\hypertarget{ref-breidt2017}{}%
Breidt, F Jay, Jean D Opsomer, and others. 2017. ``Model-Assisted Survey
Estimation with Modern Prediction Techniques.'' \emph{Statistical
Science} 32 (2). Institute of Mathematical Statistics:190--205.

\leavevmode\hypertarget{ref-carvalho2009}{}%
Carvalho, Carlos M., Nicholas G. Polson, and James G. Scott. 2009.
``Handling Sparsity via the Horseshoe.'' In \emph{Proceedings of the
Twelth International Conference on Artificial Intelligence and
Statistics}, edited by David van Dyk and Max Welling, 5:73--80.
Proceedings of Machine Learning Research. Hilton Clearwater Beach
Resort, Clearwater Beach, Florida USA: PMLR.
\url{http://proceedings.mlr.press/v5/carvalho09a.html}.

\leavevmode\hypertarget{ref-chaudhuri2011}{}%
Chaudhuri, Sanjay, and Malay Ghosh. 2011. ``Empirical Likelihood for
Small Area Estimation.'' \emph{Biometrika} 98 (2). Biometrika
Trust:473--80. \url{http://www.jstor.org/stable/23076164}.

\leavevmode\hypertarget{ref-efron1971}{}%
Efron, Bradley, and Carl Morris. 1971. ``Limiting the Risk of Bayes and
Empirical Bayes Estimators---Part I: The Bayes Case.'' \emph{Journal of
the American Statistical Association} 66 (336):807--15.
\url{https://doi.org/10.1080/01621459.1971.10482348}.

\leavevmode\hypertarget{ref-fay1979}{}%
Fay, Robert E., and Roger A. Herriot. 1979. ``Estimates of Income for
Small Places: An Application of James-Stein Procedures to Census Data.''
\emph{Journal of the American Statistical Association} 74
(366a):269--77. \url{https://doi.org/10.1080/01621459.1979.10482505}.

\leavevmode\hypertarget{ref-george2000}{}%
George, EdwardI., and Dean P. Foster. 2000. ``Calibration and Empirical
Bayes Variable Selection.'' \emph{Biometrika} 87 (4):731--47.
\url{https://doi.org/10.1093/biomet/87.4.731}.

\leavevmode\hypertarget{ref-ghitza2013}{}%
Ghitza, Yair, and Andrew Gelman. 2013. ``Deep Interactions with Mrp:
Election Turnout and Voting Patterns Among Small Electoral Subgroups.''
\emph{American Journal of Political Science} 57 (3). Wiley Online
Library:762--76.

\leavevmode\hypertarget{ref-ghosh1994small}{}%
Ghosh, Malay, and JNK Rao. 1994. ``Small Area Estimation: An
Appraisal.'' \emph{Statistical Science}. JSTOR, 55--76.

\leavevmode\hypertarget{ref-james1961}{}%
James, W., and Charles Stein. 1961. ``Estimation with Quadratic Loss.''
In \emph{Proceedings of the Fourth Berkeley Symposium on Mathematical
Statistics and Probability, Volume 1: Contributions to the Theory of
Statistics}, 361--79. Berkeley, Calif.: University of California Press.
\url{https://projecteuclid.org/euclid.bsmsp/1200512173}.

\leavevmode\hypertarget{ref-lumley2004}{}%
Lumley, Thomas. 2004. ``Analysis of Complex Survey Samples.''
\emph{Journal of Statistical Software, Articles} 9 (8):1--19.
\url{https://doi.org/10.18637/jss.v009.i08}.

\leavevmode\hypertarget{ref-mercer2014}{}%
Mercer, Laina, Jon Wakefield, Cici Chen, and Thomas Lumley. 2014. ``A
Comparison of Spatial Smoothing Methods for Small Area Estimation with
Sampling Weights.'' \emph{Spatial Statistics} 8. Elsevier:69--85.

\leavevmode\hypertarget{ref-morris1983}{}%
Morris, Carl N. 1983. ``Parametric Empirical Bayes Inference: Theory and
Applications.'' \emph{Journal of the American Statistical Association}
78 (381):47--55. \url{https://doi.org/10.1080/01621459.1983.10477920}.

\leavevmode\hypertarget{ref-pfeffermann2013}{}%
Pfeffermann, Danny, and others. 2013. ``New Important Developments in
Small Area Estimation.'' \emph{Statistical Science} 28 (1). Institute of
Mathematical Statistics:40--68.

\leavevmode\hypertarget{ref-rabe2006}{}%
Rabe-Hesketh, Sophia, and Anders Skrondal. 2006. ``Multilevel Modelling
of Complex Survey Data.'' \emph{Journal of the Royal Statistical
Society: Series A (Statistics in Society)} 169 (4). Wiley Online
Library:805--27.

\leavevmode\hypertarget{ref-raftery2012}{}%
Raftery, Adrian E, Nan Li, Hana Ševčíková, Patrick Gerland, and Gerhard
K Heilig. 2012. ``Bayesian Probabilistic Population Projections for All
Countries.'' \emph{Proceedings of the National Academy of Sciences} 109
(35). National Acad Sciences:13915--21.

\leavevmode\hypertarget{ref-rao2015}{}%
Rao, JNK, and Isabel Molina. 2015. \emph{Small Area Estimation}. John
Wiley \& Sons.

\leavevmode\hypertarget{ref-sarndal2010}{}%
Särndal, Carl-Erik. 2010. ``The Calibration Approach in Survey Theory
and Practice.'' \emph{Survey Methodology} 33 (2):99--119.

\leavevmode\hypertarget{ref-skinner2017}{}%
Skinner, Chris, and Jon Wakefield. 2017. ``Introduction to the Design
and Analysis of Complex Survey Data.'' \emph{Statist. Sci.} 32 (2). The
Institute of Mathematical Statistics:165--75.
\url{https://doi.org/10.1214/17-STS614}.

\leavevmode\hypertarget{ref-steorts2014}{}%
Steorts, Rebecca C. 2014. ``Smoothing, Clustering, and Benchmarking for
Small Area Estimation.'' \emph{arXiv Preprint arXiv:1410.7056}.

\leavevmode\hypertarget{ref-steorts2014comments}{}%
Steorts, Rebecca C, and M Dolores Ugarte. 2014. ``Comments on:`Single
and Two-Stage Cross-Sectional and Time Series Benchmarking Procedures
for Small Area Estimation'.'' \emph{Test} 23 (4). Springer:680--85.

\leavevmode\hypertarget{ref-sterba2009}{}%
Sterba, Sonya K. 2009. ``Alternative Model-Based and Design-Based
Frameworks for Inference from Samples to Populations: From Polarization
to Integration.'' \emph{Multivariate Behavioral Research} 44 (6). Taylor
\& Francis:711--40.

\leavevmode\hypertarget{ref-tibshirani1996}{}%
Tibshirani, Robert. 1996. ``Regression Shrinkage and Selection via the
Lasso.'' \emph{Journal of the Royal Statistical Society. Series B
(Methodological)} 58 (1). {[}Royal Statistical Society,
Wiley{]}:267--88. \url{http://www.jstor.org/stable/2346178}.

\leavevmode\hypertarget{ref-you2002}{}%
You, Yong, and JNK Rao. 2002. ``A Pseudo-Empirical Best Linear Unbiased
Prediction Approach to Small Area Estimation Using Survey Weights.''
\emph{Canadian Journal of Statistics} 30 (3). Wiley Online
Library:431--39.

\end{document}